\documentclass{sig-alternate-2013}

\permission{\copyright 2017 International World Wide Web Conference Committee \\ (IW3C2), published under Creative Commons CC BY 4.0 License.}
\conferenceinfo{WWW'17 Companion,}{April 3--7, 2017, Perth, Australia.}
\copyrightetc{ACM \the\acmcopyr}
\crdata{978-1-4503-4914-7/17/04. \\
http://dx.doi.org/10.1145/3041021.3054172  \\
\includegraphics{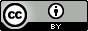}
}

\usepackage{color}
\usepackage{url}
\usepackage{verbatim} % allows multiline comments
\usepackage{multirow}
\usepackage{algorithm}
\usepackage[noend]{algorithmic}
\usepackage{xspace}
\usepackage{graphicx}
\usepackage{epsfig}
\usepackage{amsmath}
\usepackage{amssymb}
\usepackage{times}
\usepackage{float}
\usepackage{balance}
\usepackage{subcaption}
\usepackage[inline]{enumitem}

\usepackage{epigraph} % for quote

\usepackage{etoolbox}
\apptocmd{\thebibliography}{\small}{}{}

\usepackage[font=bf,belowskip=0pt,aboveskip=5pt]{caption}

\usepackage{booktabs} % for tables, bottom rule etc
\usepackage{paralist}
\renewenvironment{itemize}[1]{\begin{compactitem}#1}{\end{compactitem}}
\renewenvironment{enumerate}[1]{\begin{compactenum}#1}{\end{compactenum}}

\newcommand{\hide}[1]{}
\newcommand{\xhdr}[1]{\vspace{1.7mm}\noindent{{\bf #1.}}}

\newcommand{\eg}{\emph{e.g.}}
\newcommand{\ie}{\emph{i.e.}}

\graphicspath{{./FIG/}}

\begin{document}
% Copyright
% \setcopyright{acmcopyright}

% Tim I like this title best but perhaps the second is better to get it accepted into the track?
% \title{How competition affects user behavior: A case study of walking challenges in a mobile activity tracking application}
%\title{Gamification Increases Physical Activity: Large-scale Analysis of Walking Challenges in a Mobile Application}
\title{How Gamification Affects Physical Activity: \\ Large-scale Analysis of Walking Challenges in\\ a Mobile Application}
%Amin Changed the title from
%Competitions Increase Exercise:  Analysis of Mobile App Walking Challenges and Design Implications}

% \numberofauthors{4}
% \author{
% \alignauthor
% Ali Shameli\\
%       \affaddr{Stanford University}\\
%       \email{shameli@stanford.edu}
% %\alignauthor 
% \and
% Tim Althoff\\
%       \affaddr{Stanford University}\\
%       \email{althoff@cs.stanford.edu}
% %\alignauthor 
% \and
% Amin Saberi\\
%       \affaddr{Stanford University}\\
%       \email{saberi@stanford.edu}
% %\alignauthor 
% \and
% Jure Leskovec\\
%       \affaddr{Stanford University}\\
%       \email{jure@cs.stanford.edu}
% }

\numberofauthors{1}
\author{
\begin{tabular}{cccc}
% \mp
Ali Shameli & Tim Althoff & Amin Saberi & Jure Leskovec\\
\affaddr{Stanford University} & \affaddr{Stanford University} & \affaddr{Stanford University} & \affaddr{Stanford University}\\
\email{shameli@stanford.edu} & \email{althoff@cs.stanford.edu} & \email{saberi@stanford.edu} & \email{jure@cs.stanford.edu}
\end{tabular}
}

\maketitle

\begin{abstract}
% !TEX root = paper-competition.tex

%While physical activity helps with maintaining a healthy weight and reduces the risk for several chronic diseases, most people around the world do not get the recommended amounts of activity. One of the methods that has been successful in increasing physical activity is gamification. However, our understanding of contributing mechanisms including competition is limited. 

Gamification represents an effective way to incentivize user behavior across a number of computing applications. However, despite the fact that physical activity is essential for a healthy lifestyle, surprisingly little is known about how gamification and in particular competitions shape human physical activity.

Here we study how competitions affect physical activity. We focus on walking challenges in a mobile activity tracking application where multiple users compete over who takes the most steps over a predefined number of days. We synthesize our findings in a series of game and app design implications. In particular, we analyze nearly 2,500 physical activity competitions over a period of one year capturing more than 800,000 person days of activity tracking. We observe that during walking competitions, the average user increases physical activity by 23\%.  Furthermore, there are large increases in activity for both men and women across all ages, and weight status, and even for users that were previously fairly inactive. We also find that the composition of participants greatly affects the dynamics of the game. In particular, if highly unequal participants get matched to each other, then competition suffers and the overall effect on the physical activity drops significantly. Furthermore, competitions with an equal mix of both men and women are more effective in increasing the level of activities. We leverage these insights to develop a statistical model to predict whether or not a competition will be particularly engaging with significant accuracy. Our models can serve as a guideline to help design more engaging competitions that lead to most beneficial behavioral changes.

\end{abstract}

% % \vspace{1mm}
% \noindent {\bf Categories and Subject Descriptors:} H.2.8 {\bf
% [Database Management]}: Database applications---{\it Data mining}

% % \noindent {\bf General Terms:} Algorithms; Experimentation.

% \noindent {\bf Keywords:} 

\section{Introduction}
% !TEX root = paper-competition.tex

% \epigraph{Those who think they have not time for bodily exercise will sooner or later have to find time for illness.}{Edward Stanley, Earl of Derby, 20 December 1873}

% TIM: cut for space 
% \vspace{-4mm}
% \epigraph{If exercise could be packed in a pill,
% it would be the single most widely
% prescribed and beneficial medicine
% in the nation.}{Robert N. Butler. Director, Nat. Institute on Aging} %  M.D.;
% \vspace{-3mm}

Physical activity is critical to human health~\cite{WHO2010parecommendation}.  People who are physically active tend to live longer, have lower risk of several diseases including heart disease, stroke, Type 2 diabetes, depression, and some types of cancer, and are more likely to maintain a healthy weight (\eg, \cite{miles2007physical,sparling2000promoting,althoff2016quantifying}).
However, only 21\% of US adults meet official physical activity guidelines~\cite{cdc2014pafacts,us2008physical} (at least 150 minutes a week of physical activity for adults), and less than 30\% of US high school students get at least 60 minutes of physical activity every day~\cite{cdc2014pafacts}. 
%Inactivity is costing lives and money. 
It is estimated that physical inactivity contributes to 5.3 million deaths per year worldwide~\cite{lee2012effect} and that it is responsible for a worldwide economic burden of \$67.5 billion through health-care expenditure and productivity losses~\cite{ding2016economic}. 

%Therefore, any efforts aimed at increasing physical activity across the population could make a significant impact on the public health and 
Given huge potential to improve public health, many interventions and small-scale studies have been designed towards increasing physical activity across the population (\eg, \cite{dishman1985determinants, dobbins2009school, marshall2004challenges, reis2016scaling,sallis1998environmental,salmon2007promoting}).
Unfortunately, many of these interventions are deemed either ineffective~\cite{dobbins2009school,salmon2007promoting} or are limited in that they only reach small populations~\cite{dishman1985determinants,marshall2004challenges}.
Recently, however, gamification techniques have become widely adopted and have been very impactful in obtaining behavioral outcomes~\cite{hamari2014does}. Successful examples for incentivizing physical activity through so-called {\em exergames}~\cite{gobel2010serious,sinclair2007considerations,staiano2011exergames} include in-game avatars~\cite{lin2006fish} and geo-centric games such as Pok\'emon Go~\cite{althoff2016pokemon}.
%While several studies have emphasized the importance of social influence to increase physical activity~\cite{althoff2017onlineactions,consolvo2006design}, we do not fully understand the various social influence mechanisms and associated design implications. 
%One example of such mechanisms that could potentially lead to increased physical activity are online competitions. 
However, basic gamification mechanisms such as competitions and challenges have been relatively poorly explored and understood. While competitiveness is found to be associated with greater enjoyment~\cite{frederick2003competition}, there have been no quantitative studies whether and how such competitions affect physical activity. 
Given the proliferation of mobile devices and health and activity tracking applications, effective and engaging competitions that increase physical activity have a huge potential to achieve population-wide improvements in public health and decrease in risk of various chronic diseases. %, and lead to longer life spans.

Here we study how various game design elements used by mobile health apps encourage exercise, fitness, and essentially weight loss. We analyze the effect of competitions on increasing the level of physical activity of participants. 
We study user physical behavior as captured by the Azumio Argus activity tracking app. The application allows users to create and engage in ad-hoc games that last from one to seven days and include an arbitrary number of  participants. Participants then compete over who takes the highest total number of steps over the predefined duration of the competition.  
The dataset obtained from Argus contains nearly 2,500 physical activity competitions over a period of one year capturing over 800,000 person days of in-competition activity tracking. For each user we have a record of their daily physical activity, competition participation, as well as their demographic data (gender, age, height, and weight). This constitutes the largest studied dataset of physical activity competitions to date.

% Results
% \todo{Update all numbers below. Currently they are inaccurate.} Ali updated the numbers. 
%\todo{this will need to be revisited based on the results sections :)}

We analyze how participation in the game impacts the activity of participants during the time of the competition compared with their baseline activity levels.
We find that during walking competitions, the average user increases their physical activity by 23\%. 
Furthermore, we show that there are large increases in activity for both men and women across all ages, weight status, and baseline activity levels.
In fact, we find the largest increases for users that were previously fairly inactive who exhibit an average increase of more than 2,500 steps per day throughout the challenge.
Increases of this magnitude -- if sustained over time -- could lead to significant improvements in participants' health outcomes.

Then, we turn our attention to quantifying how much effort it takes to win a competition and find that winners increase their activity by 40-60\% while the last-ranked users are on average less active than they were before.  We also observe that the winner's effort increases in competitions with more participants. 

We also find that the composition of participants greatly affects the dynamics of the competition leading to important design implications for exergames and mobile health applications:   
%In particular, if highly unequal participants get matched to compete, then the overall dynamics of the competition suffers and the overall effect of the competition on the physical activity drops significantly. Furthermore, competitions with a balanced mix of both men and women are more effective in increasing activity.
% \todo{
\begin{enumerate}
  \item Competitions lead to increases in physical activity and constitute a viable design element able to reach a broad user base across a wide variety of user demographics. 
  \item Competing participants should have similar pre-com\-peti\-tion activity levels. Otherwise the effect of the competition on physical activity drops significantly.
  \item Competitions should have a balanced mix of both men and women.
  \item Competitions should ideally include some participants who have previously increased their activity in response to competitions to encourage the other participants.
\end{enumerate}
% } %end todo 

We leverage these insights in a statistical model that predicts whether or not a competition will be particularly engaging to  the participants. Our model can serve as a guideline to help group participants into competitions that are more competitive and thus lead to highest behavioral changes. Our approach can be potentially used across a variety of mobile health applications and games to recommend evenly-matched competitions to users which are most likely to be more active.

%\amin{ Jure: please feel free to expand this section and include a few other observations including: 1- the second rank participant increases his or her activity but not as much as the first. 2- in competitions with 5 participants, the third \& fourth ranked did not change their activity much 3- observations from section 6} 

% Contributions
%In summary, our contributions are:
%\begin{enumerate}
  %\item ...
%\end{enumerate}

% Implications
%Our work has implications for the design of competitions in online and mobile health applications and for the development of physical activity interventions. 

\section{Related Work}
% !TEX root = paper-competition.tex

% Our work relates to N lines of research.

\noindent Next we survey related work and discuss our work in context. 

\xhdr{Physical activity}
The link between physical activity and improved health outcomes has been well-established (\eg,~\cite{ding2016economic,lee2012effect,miles2007physical,sparling2000promoting,WHO2010parecommendation,althoff2016quantifying}).  
At the same time, only a small fraction of people in developed countries meet official physical activity guidelines~\cite{cdc2014pafacts,us2008physical}.
While, many interventions are aimed at increasing physical activity (\eg,~\cite{dishman1985determinants,dobbins2009school,marshall2004challenges, reis2016scaling,sallis1998environmental,salmon2007promoting}), 
many of them seem ineffective~\cite{dobbins2009school,salmon2007promoting} or were only reaching already active populations instead~\cite{dishman1985determinants,marshall2004challenges}.

\xhdr{Measurement of physical activity}
Consumer wearable devices and smart phones are becoming more prevalent in the general population and could enable a better understanding of real-world health behaviors and physical activity and how to best support and encourage healthier behaviors~\cite{hayden2016mobile,servick2015mind,althoff2017harnessing}.
However, few research studies to date have harnessed data obtained from consumer wearables to study physical activity~\cite{althoff2017onlineactions,althoff2016pokemon}. 
Medical studies have examined accelerometer-defined activity (\eg,~\cite{troiano2008physical,tudor2004many}), but much of the social media research related to human health has relied on self-reports and proxy measures (\eg, \cite{de2016discovering,de2016characterizing,mejova2015foodporn}), which have been found to be severely biased~\cite{Tucker2011selfreport}. In contrast, we use objective physical activity measurements from smart phone accelerometers.

\xhdr{Incentivizing user behavior}
Studies have found that use of pedometers and activity trackers for self-monitoring can help increase activity~\cite{thorup2016cardiac,wang2016mobile} but other studies have reported mixed results~\cite{wang2015wearable}.
Beyond enabling self-monitoring, encouraging additional activity through reminders led to increased activity only for the first week after the intervention and did not lead to any significant changes after six weeks in a randomized controlled trial~\cite{wang2015wearable}.
However, in online domains gamification has been successful in changing user behavior~\cite{hamari2014does,souza2016evaluating}. For example, badges increase engagement in question answering sites~\cite{anderson2013steering},  online courses~\cite{anderson2014engaging}, and pro-social behaviors~\cite{althoff2014howtoaskforafavor}.
To encourage healthy behavior, researchers have studied the design of ``exergames''~\cite{gobel2010serious,sinclair2007considerations,staiano2011exergames}, video games combined with exercise activity~\cite{lin2006fish}, and location-based games where game play progresses through the physical environment~\cite{althoff2016pokemon,avouris2012review}.
Furthermore, social networks can also modify human behavior through peer influence. 
For example, researchers have highlighted the importance of facilitating social influence to encourage more exercise~\cite{consolvo2006design} and have found that sharing exercise activity through social networks has positive long-term effects on physical activity levels~\cite{althoff2017onlineactions}. %from social network connections and allowing for sharing of activity and progress among users
Furthermore, researchers have studied social interactions to better understand how people could be most supportive of others~\cite{althoff2016counseling,althoff2014howtoaskforafavor}.

% competion

Prior research also studied competitions and competitive behavior. Competitions can improve behavioral outcomes, for example in the running speed of children in short-distance races~\cite{gneezy2004gender}. Men are more likely to embrace competitive formats than women~\cite{niederle2007women}, and while competitiveness is associated with greater enjoyment~\cite{frederick2003competition}, there have been no quantitative studies of how competitions affect physical activity. 

\xhdr{This work}
Our work extends the existing literature on incentivizing healthy user behavior and exergame design implications by studying effects of competitions on physical activity. 
Unlike badges and simple activity tracking, competitions allow users to compete directly with each other. 
We use a large dataset of online competitions within a mobile activity tracking application in conjunction with objective measures of physical activity based on smart phone accelerometers. Our work has implication for the large number of mobile and web health applications using competitions to improve user engagement.

\section{Dataset Description}
% !TEX root = paper-competition.tex

We use a dataset of competitions within the Argus smartphone app by Azumio which allows users to track their daily activities.
Competitions run for 1, 3, 5 or 7 days and can have any number of participants (who may or may not know each other outside the application). 
In this paper, we focus on the longest competitions running over seven days from Monday through Sunday.
This ensures that each competition includes the exact same number of each weekday (one) and that any findings are not confounded by differences in weekday versus weekend activity. 
Furthermore, we restrict analysis to competitions with at least three participants. 
We use a dataset of 3,637 users in 2,432 competitions satisfying these constraints.
These competitions occurred over a time period of 10 months and included a total of 535 million steps over 70,413 person days.
All participants had used the activity tracking app prior to the start of their first competitions. Therefore, any increases in activity during competition periods are not an effect of self-monitoring. 
Table~\ref{tab:dataset_statistics} further summarizes the dataset and shows that the distribution of age, gender, and weight is fairly representative of the overall population in many developed countries. 
For example, the median age is 34 years (official estimate in United States is 37 years) and the data is evenly split between males and females.
Furthermore, over 34.4\% of users are overweight and 18.8\% of users are obese highlighting that not only healthy users participate in the walking competitions.
Users that have participated in at least one competition take
about 6164 daily steps on average outside of competitions.
Compare this to 5926 for users that have never participated in a competition before.
This shows that there is a slight selection effect; that is, competitions seem particularly attractive to those users with slightly elevated activity levels.

Physical activity levels are quantified using the number of steps in each day objectively measured through smartphone accelerometers (as done in many research studies~\cite{althoff2017onlineactions,althoff2016pokemon,Bassett2010,tudor2004many}). 
Objective measures are critical as commonly used self-reports of physical activity can be extremely biased with Tucker et al. reporting overestimates of up to 700\%~\cite{Tucker2011selfreport}.

Our dataset is the largest dataset of physical activity competitions studied to date.
It has three key properties:
(1) It covers a large number of competitions covering a diverse population of participants in terms of age, gender, weight status, and activity level.  
(2) The activity levels of the users are measured objectively.
(3) The activity levels of individuals are measured before and during the competition offering a baseline level for measuring the effect of competition. 
Therefore, this dataset uniquely enables the study of how online competitions affects offline physical activity.

\begin{table}[t]
\centering
 \small % \footnotesize
\resizebox{.95\columnwidth}{!}{%
\begin{tabular}{ll}
  %\hline
  \toprule
  % \textbf{Dataset Statistics} &  \\
  % \midrule
  \# 7 day competitions w. at least 3 particip. & 2,432 \\
  \# total users in competitions & 3,637 \\
  Observation period & April 2015 -- Jan. 2016\\
  \# days of steps tracking (within competition) & 70,413\\
  \# days of steps tracking (outside competition) & 817,666\\
  \# total steps tracked (inside competition) & 535 million\\ % 535,843,717
  Median age & 34 years\\
  \% users female & 51\%\\%46.13\%
  \% underweight (BMI < 18.5) & 3.1\%\\
  \% normal weight (18.5 $\leq$ BMI < 25) & 43.7\%\\
  \% overweight (25 $\leq$ BMI < 30) & 34.4\%\\
  \% obese (30 $\leq$ BMI) & 18.8\%\\
  Avg. daily steps outside competition  & \\
  for competition users & 6,164\\
  Avg. daily steps outside competition & \\
  for non-competition users & 5,926\\
  \bottomrule
 \end{tabular} }
 % \vspace{-3mm}
 \caption{
 Dataset statistics. 
 BMI refers to body mass index.
 }
 \vspace{-2mm}
 \label{tab:dataset_statistics}
 \end{table}

\section{Does Competition Increase\\Activity?}
% !TEX root = paper-competition.tex

\begin{figure}[t!]
\centering
%\begin{minipage}{.47\columnwidth}
  \centering
  \includegraphics[width=0.80\columnwidth]{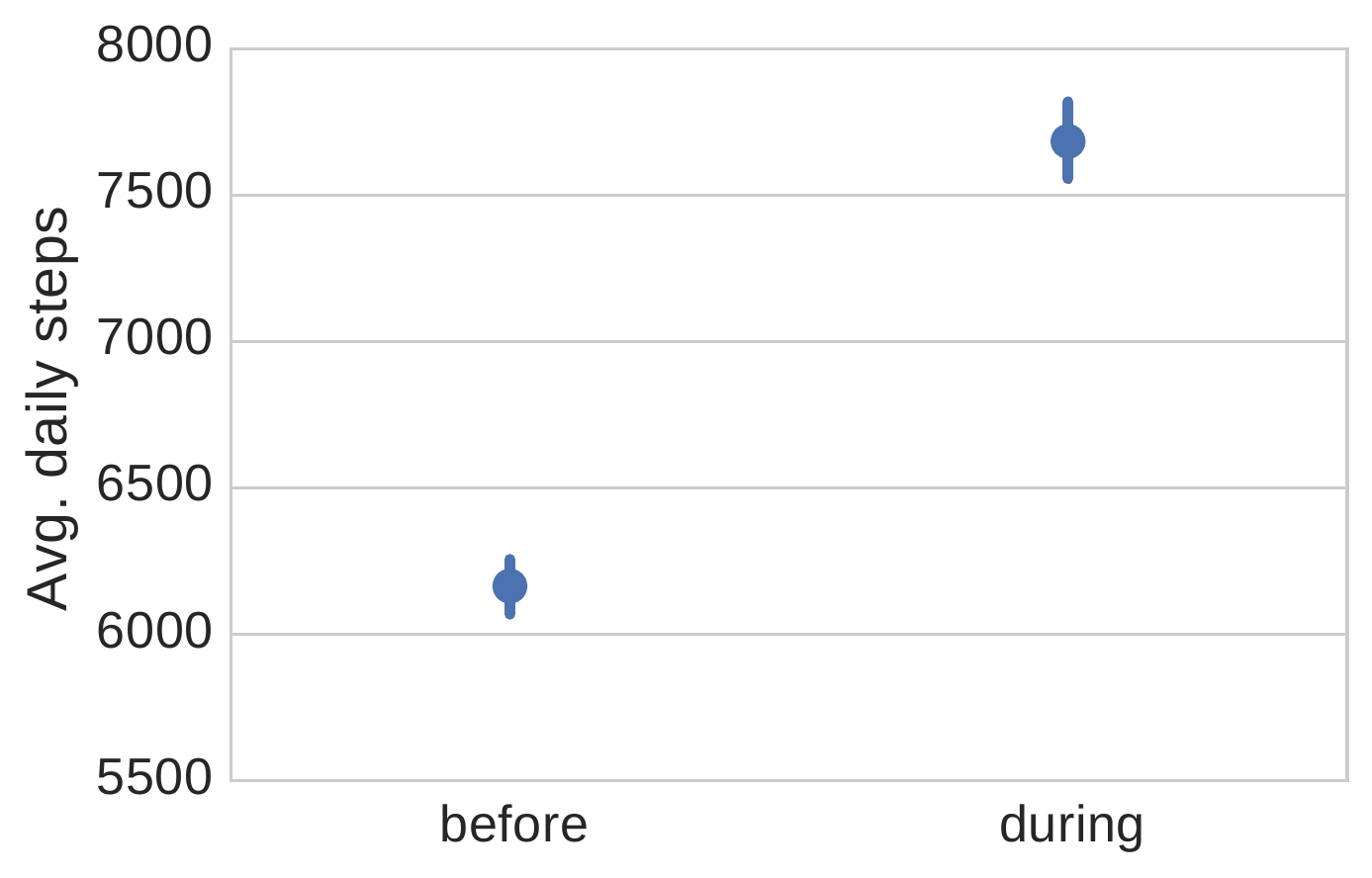}
  %\includegraphics[width=0.35\columnwidth]{FIG/WWW-stepsperday.pdf}
 %    \vspace{-2mm}
  \captionof{figure}{Average number of steps taken in the past when not participating in a competition vs. the average number of steps taken in a competition.
  We find that users take significantly more steps during competitions. 
  Error bars in all figures correspond to 95\% confidence intervals of the corresponding mean estimates.}
%     \vspace{-3mm}
%\end{minipage}%
  \label{fig:WWW-before-during}
\end{figure}

The first question we set to answer is whether competitions tend to increase user activity. Given the heterogeneity across users, we compare the physical activity of a user to his or her own baseline activity level. In other words, we measure the activity as the average number of steps per day over the duration of the competition, and compare that to the average daily activity of the user while not participating in a competition. 

\xhdr{Competitions lead to increased physical activity of participants}
We make several interesting observations (Figure~\ref{fig:WWW-before-during}). First, we notice that the average daily number of steps of a person that has ever participated in a competition is 6,164, which is about 200 steps more than an average user. This indicates that there is some selection effect and that  more active users tend to participate in physical activity competitions. Second, we observe that during the competition the overall average activity increases to more than 7,500 steps, which is a 1,400 steps per day increase over the out-of-competition activity baseline. This means that participating in a competition leads to  23\% average increase in the physical activity.

\xhdr{Discussion} 
%We note that there is a selection effect as users self-select themselves into competitions. 
%Since we study observational data, we cannot remove this effect, for example by randomizing users into competitions.
%Identifying causal effects from observational data is fundamentally challenging. 
To check for the robustness of our finding we also analyzed the differences between the group of users participaing in the competitions against those that do not participate in competitions. We found that both groups are very similar in terms of age, gender, and weight status. Furthermore, we also found that non-competition users were only slighly less active than competition users outside of competitions (5926 vs. 6164 average daily steps). On top of this we observed that the activity increase is about the same for users who start their own competitions and for users who accept an invitation to join an existing competition. All these findings suggest that insights derived from the above analyses are robust and generalize across user groups.

%However, for the 
%We emphasize that we find within user differences in longitudinal data supporting our conclusions.
%We cannot make claims about users that did not choose to join a competition. 
%Users joining competitions might be more motivated than other users.  However, w

% numbers backing up that groups are similar
% not participated:
% age mean: 35
% male: 24%
% underweight: 3%
% normal: 45%
% overweight: 32%
% obese class 1: 13%
% obese class 2: 5%

% participated:
% age mean: 34
% male: 28%
% underweight: 3%
% normal: 42%
% overweight: 33%
% obese class 1: 14%
% obese class 2: 5%

\begin{figure*}[ht!]
  \centering
    \begin{subfigure}[b]{0.5\columnwidth}
        \includegraphics[width=\textwidth]{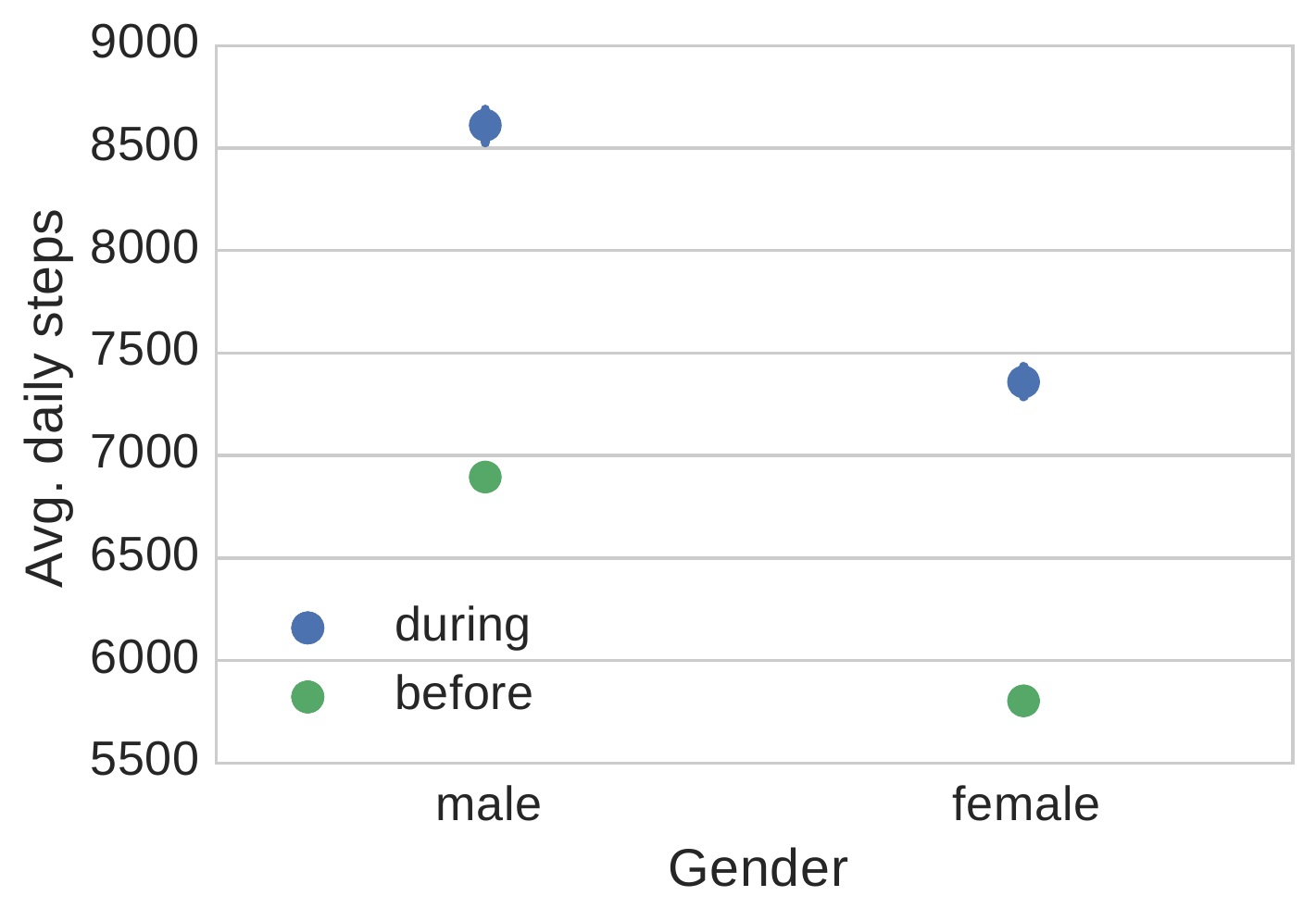}
        \caption{Gender}
        \label{fig:WWW-before-during-gender}
    \end{subfigure}
    \begin{subfigure}[b]{0.5\columnwidth}
        \includegraphics[width=\textwidth]{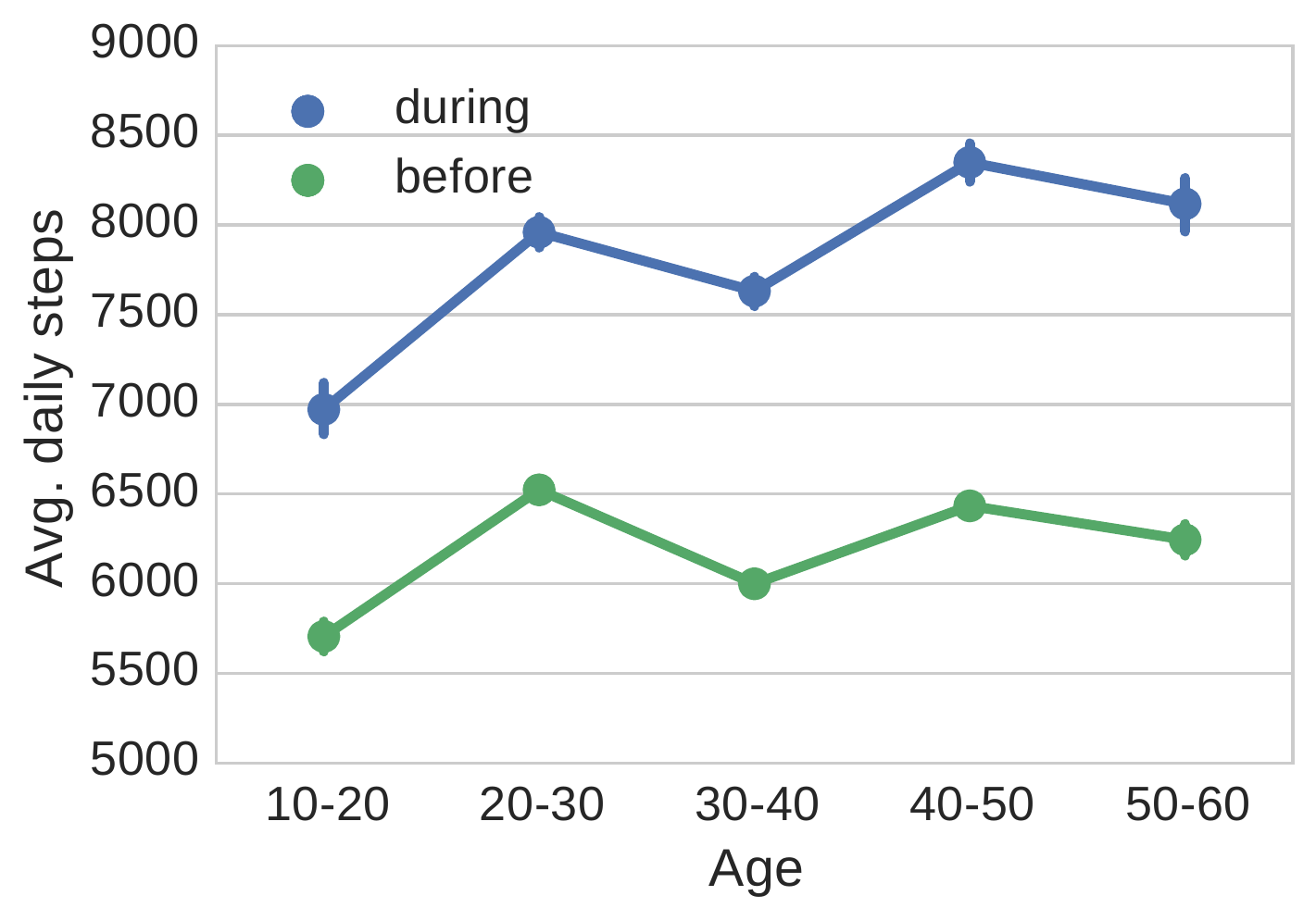}
        \caption{Age}
        \label{fig:WWW-before-during-age}
    \end{subfigure}
    \begin{subfigure}[b]{0.5\columnwidth}
        \includegraphics[width=\textwidth]{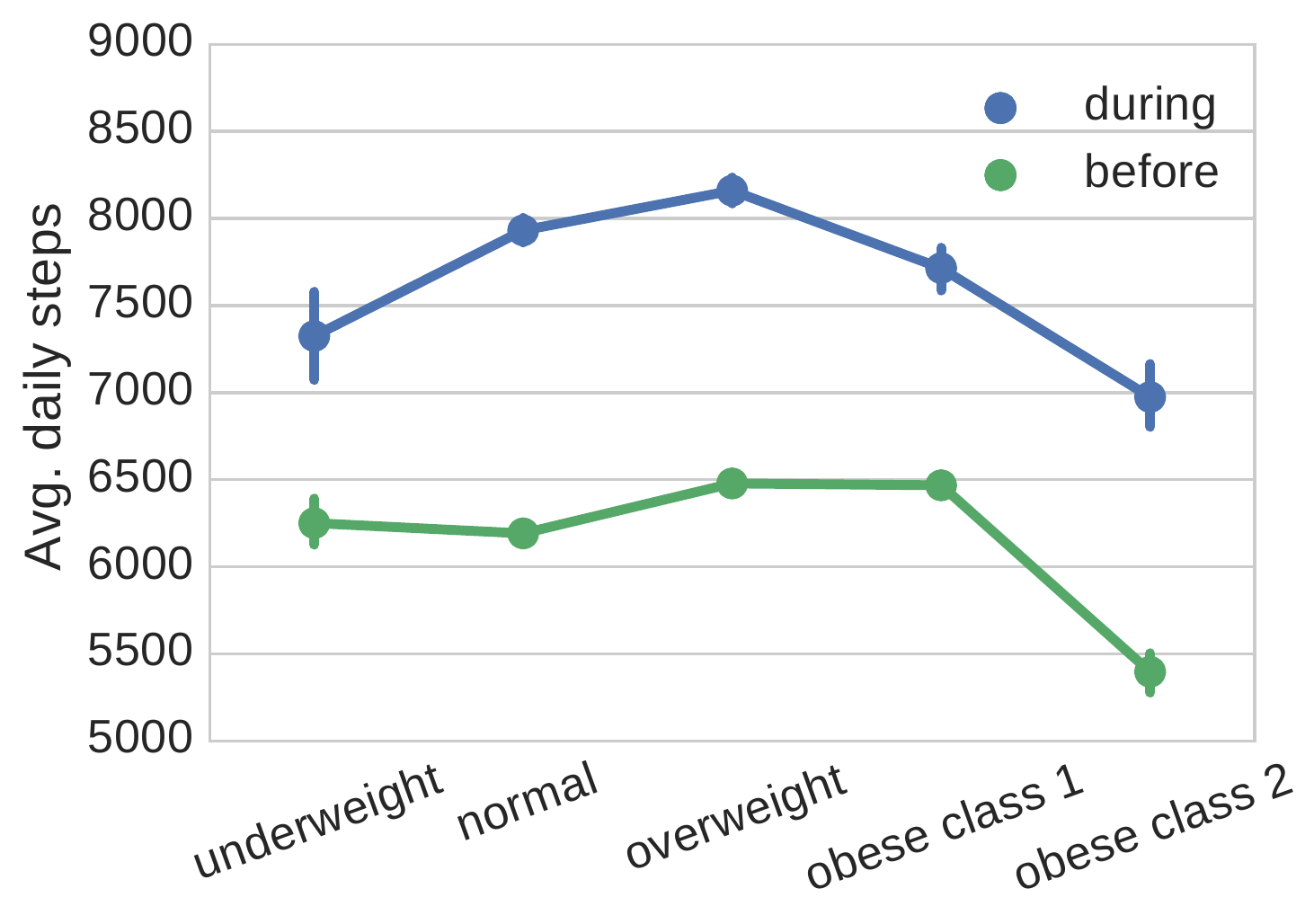}
        \caption{BMI}
        \label{fig:WWW-before-during-bmi}
    \end{subfigure}
    \begin{subfigure}[b]{0.5\columnwidth}
        \includegraphics[width=\textwidth]{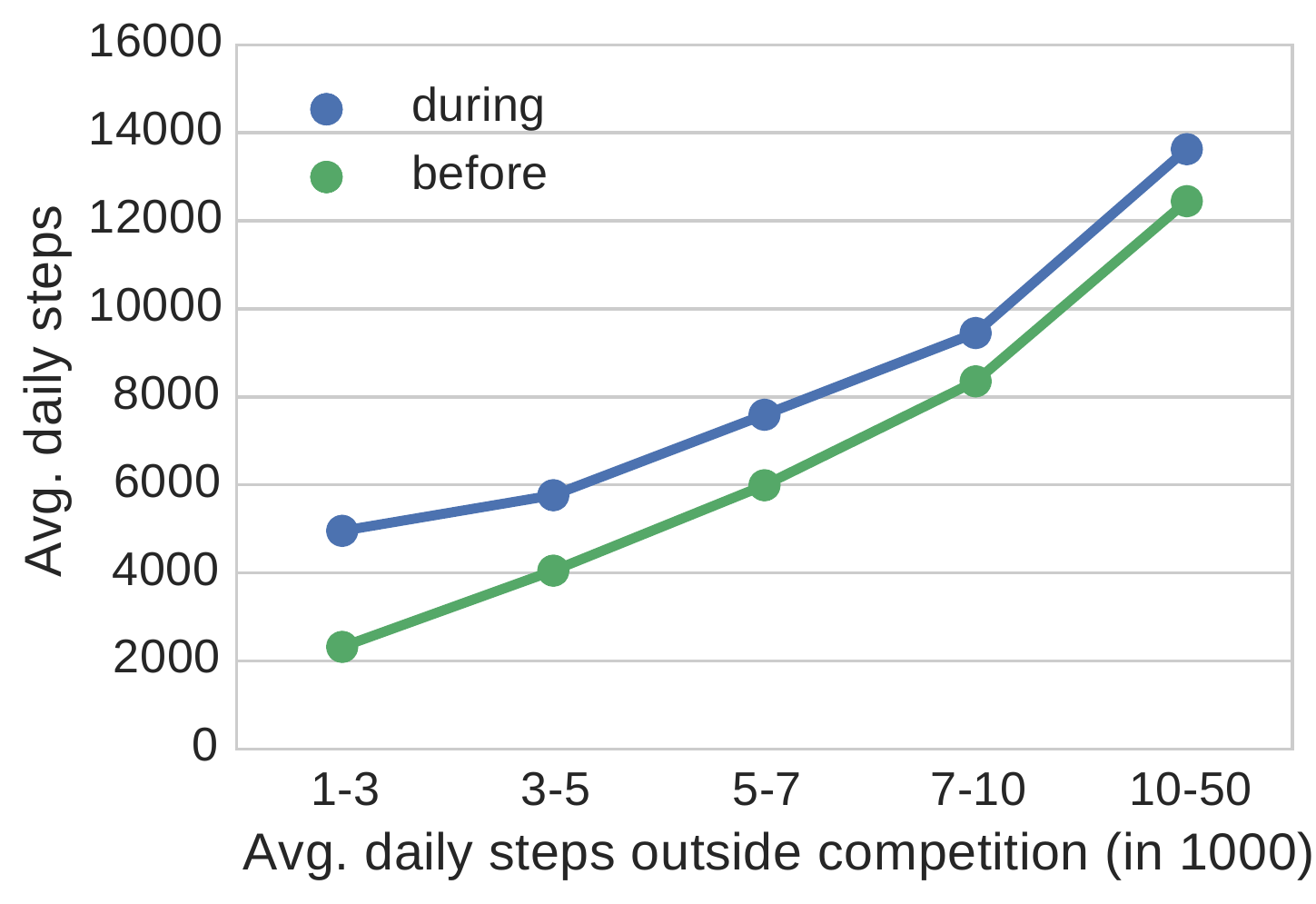}
        \caption{Previous Activity}
        \label{fig:WWW-before-during-activity}
    \end{subfigure}
    \caption{The average number of steps taken by users categorized by gender, age, BMI, and previous activity level. The green points represent the average number of steps when not participating in a competition and the blue points represent the average number of steps during competitions.}
    \label{fig:WWW-before-during-comp}
\end{figure*}

\subsection{Who is changing how much?}

Our analysis  showed that participants tend to increase their activity during competitions, which demonstrates that competitions may be an effective way for people to increase their physical activity. However, note that this increase is observed in average over all participants. The person who wins the competition may increase their activity  significantly while the person who is last in the competition may not increase their activity at all. Furthermore, demographic indicators, like gender, age, and BMI (body mass index) affect physical activity and thus 
it is not clear how these factors modulate in-competition activity of individual participants.
%\tim{finish sentence}

In order to understand which users increase their activity the most during the competitions we measure the activity before and during the competition for different demographic groups. Figure~\ref{fig:WWW-before-during-comp} shows separate plots for gender, age, BMI, and baseline activity level. We make several observations.

\xhdr{Gender}
Consistent with previous literature~\cite{Bassett2010}, Figure~\ref{fig:WWW-before-during-gender} shows that men tend to be more active than women on average. This is also true during the competitions where men increase their activity from just below 6,900 steps per day to about 8,500 steps per day. Similarly, women increase their activity from 5,800 steps per day to about 7,100 steps per day. Interestingly, in both cases we observe the same relative increase in activity. Both men and women tend to increase their activity by 23\% during competitions.

% activity decreases with increasing age\cite{Bauman2012,Bassett2010,Troiano2008,Hallal2012}  and is lower in females than in males\cite{Bassett2010,Troiano2008,Bauman2012,Hallal2012}. 

\xhdr{Age} Examining physical activity levels as a function of age (Figure~\ref{fig:WWW-before-during-age}), we find that the difference of activity levels across different age groups is rather minimal -- people below age 20 are the least active group with 5,700 steps per day (outside competitions), while the age 20-30 group is the most active with 6,500 steps per day. Furthermore, we do not observe an expected decrease in physical activity as participants get older~\cite{Bassett2010}. Surprisingly, even the group of 50-60 year old users take  nearly 6,200 steps per day. We attribute this observation to a selection effect as a small but physically very active fraction of this age group may be participating in competitions.
Perhaps more interestingly we observe that regardless of the age group, physical activity during competitions tends to increase for about 1,400 steps. The increase in physical activity is robust. For example, for the age group 10-20 the increase is 1,300 steps per day (22\%), and then increases nearly linearly so that the age group 50-60 exhibits an increase of 1,800 steps per day (28\%). 

\xhdr{Body mass index (BMI)} Next, we examine how competitions affect people with different body mass index (BMI). Using self-reported height and weight we compute each participant's BMI and group them into five groups: underweight (15-18.5 BMI), normal (18.5-25 BMI), overweight (25-30 BMI), and then two groups of obese people (30-35 BMI and 35-40 BMI, respectively). In Figure~\ref{fig:WWW-before-during-bmi} we observe that baseline physical activity is relatively constant for the first four groups and lower for only the most obese group (5,800 vs. 6,300 steps per day). Interestingly, however, the increase in physical activity is consistent across all the BMI groups. Underweight people increase their activity during competitions the least (1,000 steps, 18\%), while the increase is the highest for normal-weight and overweight people (1,700 steps, 28\%).

\xhdr{Baseline activity level} Last, we examine how previous activity modulates activity inside competitions. We find that competitions lead to increased physical activity regardless of baseline activity rate (Figure~\ref{fig:WWW-before-during-activity}). Moreover,  low activity people tend to be most affected by the competitions---they increase their activity the most (to 4,700 steps per day; 103\% increase) and both the total increase in activity as well as the relative increase tend to decay with the activity level before competitions. For example, the increase due to competitions is only 1,000 steps per day (9\%) for people with over 10,000 average daily steps.

To conclude, we observe that competitions have significant and measurable effect on the physical activity of participants. Regardless of the gender, age group, or the body mass index, we observe a robust increase of about 1,400 steps (about 23\%) per day in the physical activity of participants.
If sustained, large increases like this would have a significant positive effect on the health of the participants~\cite{ding2016economic,dwyer2015objectively,lee2012effect,miles2007physical,sparling2000promoting,WHO2010parecommendation}.

%Figures~\ref{fig:WWW-before-during-gender}, \ref{fig:WWW-before-during-bmi}, \ref{fig:WWW-before-during-age}
%then by age, gender, bmi, previous activity level

\section{What Does It Take To Win A\\Competition?}
% !TEX root = paper-competition.tex

%figure 1 (final)
%Fig 10/15 and 10/30-33 in one plot
%maybe fig 10/39

%Figures \ref{fig:WWW-final-beforeduringafter_lineplot}, \ref{fig:WWW-stepsperday}, \ref{fig:WWW_steps_per_counts}, \ref{fig:WWW_delta_steps_per_counts}, \ref{fig:WWW-steps-mean-perday_5participants_cumrank}

\begin{figure}[t!]
  \centering
  \includegraphics[width=1\columnwidth]{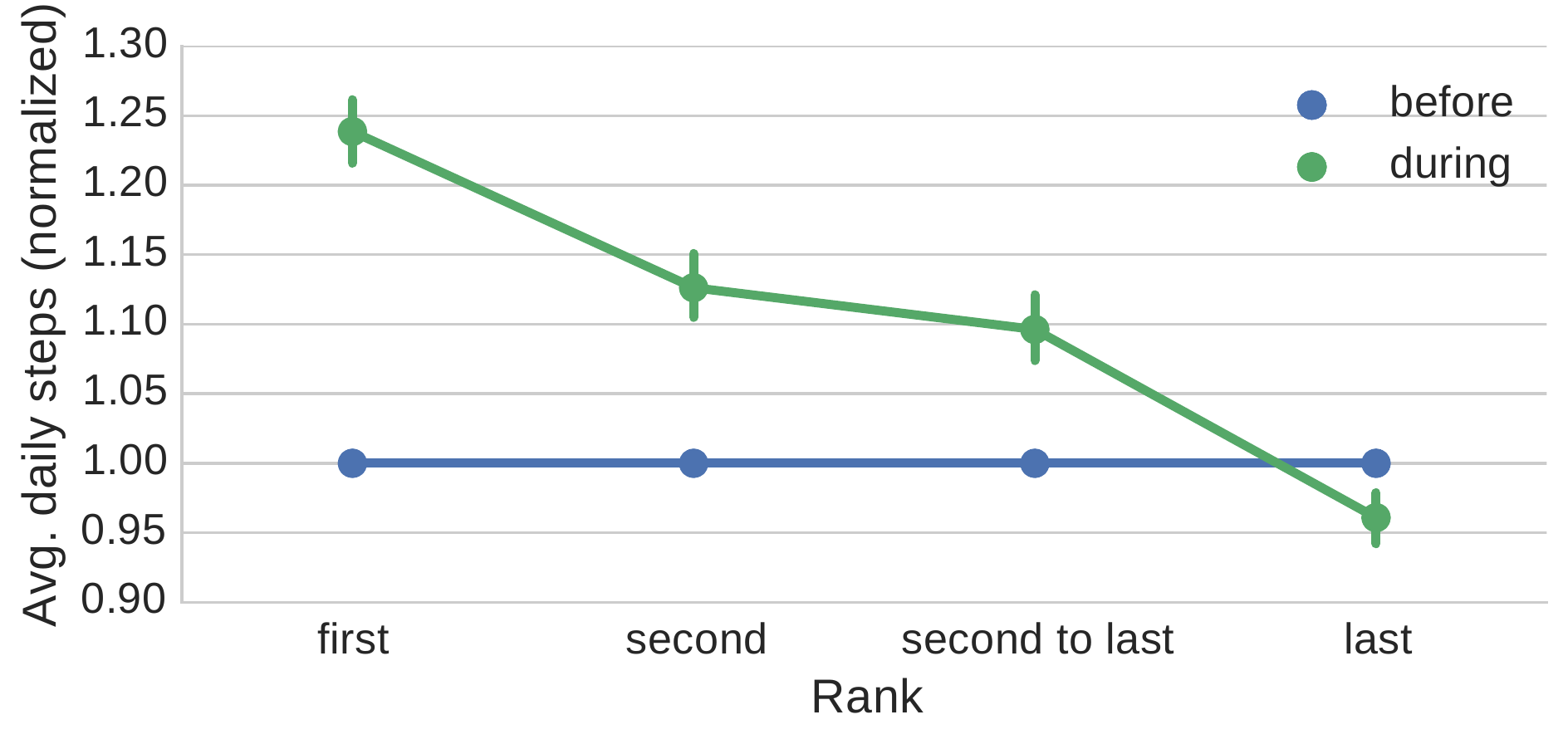}
 %    \vspace{-5mm}
  \captionof{figure}{Avg. relative number of steps taken one week before and during a competition for the two top and bottom users.}
%     \vspace{-3mm}
  \label{fig:WWW-final-beforeduringafter_lineplot} 
\end{figure}

In the previous section we observed remarkably strong effect of competitions on physical activity of participants. While the increase is stable across gender, age and weight groups, in the end there can only be one winner of the competition. So, in this section we investigate how much activity is needed in order to win a competition.

\xhdr{It takes 25\% more steps per day to win}
We analyze all competitions with duration of seven days that have at least three participants. 
This considers the following final placements: 
% This means that each competition has four well defined final placements:
 first, second, second to last, and last. 
We measure how much does a participant need to increase their physical activity (measured in the number of steps per day) compared to the baseline activity which is their average number of steps one week before the competion.
%when they are not in a competition. 
Figure~\ref{fig:WWW-final-beforeduringafter_lineplot} shows the relative change in the number of steps as a function of the final placement of the participant.

We observe that the winner of the competition increases his or her activity for 25\% over their baseline activity, while the second person increases it for only 13\%, second to last person still increases their activity over the baseline for about 10\%, while the last person actually drops their activity below their baseline activity (4\% drop). We conclude that winners increase their activity the most, followed by the mid-placed people, while the last person slightly decrease their activity. Regardless of this, the overall average activity across all participants in the competition is still higher than outside the competition.

\begin{figure}[t!]
  \centering
    \begin{subfigure}[b]{0.49\columnwidth} % .48
        \includegraphics[width=\textwidth]{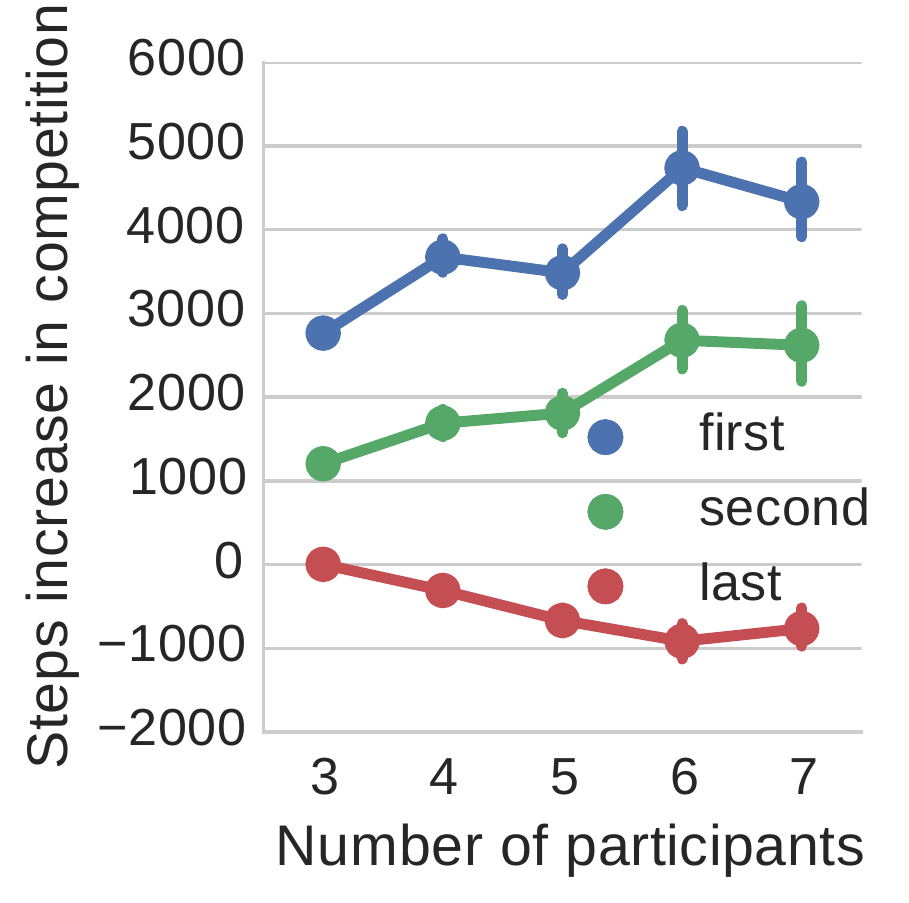}
        \caption{Steps per day}
        \label{fig:WWW_steps_per_counts}
    \end{subfigure}
    \begin{subfigure}[b]{0.49\columnwidth}
        \includegraphics[width=\textwidth]{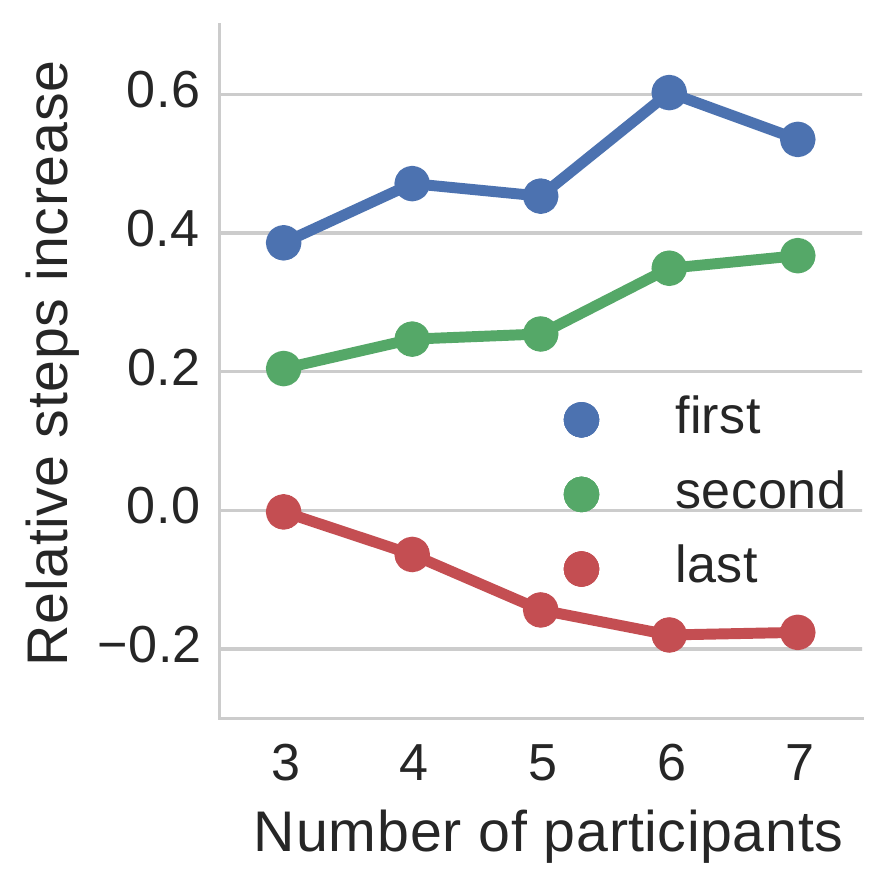}
        \caption{Steps delta}
        \label{fig:WWW_delta_steps_rel_per_counts}
    \end{subfigure}
    \caption{(a) Average number of steps taken by users by the number of participants in the competition. (b) Average increase in steps compared to previous outside competition activity by the number of participants in the competition.
    %\tim{I suggest using only horizontal gridlines in b to make it consistent with all the other plots.} \jure{Make the style of the plot (b) match the style of plot (a)!} \ali{Done!}
    }
%       \vspace{-3mm}
    \label{fig:WWW-steps_per_counts_comp}
\end{figure}

\xhdr{Winner's effort increases as there are more competitors}
We also examine how does the winner's effort increase as there are more competitors in a competition. Figure~\ref{fig:WWW_steps_per_counts} plots the absolute number of steps per day of the winner as a function of participants in the competition, while Figure~\ref{fig:WWW_delta_steps_rel_per_counts} plots the increase in the number of steps when compared to the out-of-the-competition baseline. Each separate curve plots the activity of the top, second, and last placed participants.

In Figure~\ref{fig:WWW_steps_per_counts} we observe that the absolute activity of the winner and the second-placed participant increase with the competition size while the activity of the last-placed person actually decreases. Examining data in Figure~\ref{fig:WWW_delta_steps_rel_per_counts} we find that with each additional participant in the competition the final winner of the competition increases their activity for 420 steps, while the effect on the second-placed participant is about 100 steps smaller (320 steps per addit. participant).

\begin{figure*}[t]
  \centering
  \includegraphics[width=1\textwidth]{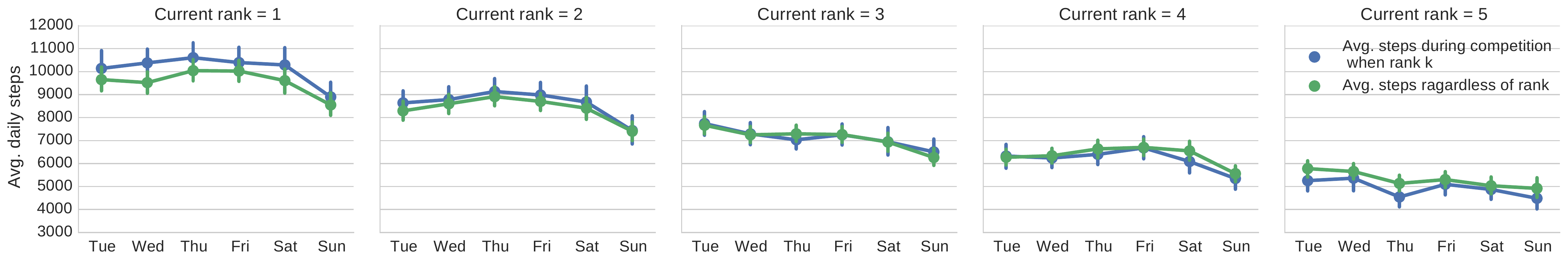}
%     \vspace{-4mm}
  \captionof{figure}{Average number of steps taken by users with specific current ranks in the competition over different days of the week for 7-day competitions with 5 participants. 
  %\jure{Fix green legend. It is currnetly not accurate. Blue: ``Avg. steps during competition when rank  k'', Green: ``Avg. steps regardless of rank''}
  %\tim{is this the correct legend? I thought it was both during competition? } \ali{Done!}
  }
     % \vspace{-10mm}
  \label{fig:WWW-steps-mean-perday_5participants_cumrank}
\end{figure*}

\xhdr{Activity over time}
Last, we also examine how participant activity changes over time. We examine only competitions with exactly 5 participants to control for competition size. After every day of the competition we compute the current position of every participant and ask: given the participant was ranked $k$ ``yesterday'', how many steps are they going to do today?

Figure~\ref{fig:WWW-steps-mean-perday_5participants_cumrank} plots the average daily steps ``today'' for a person who was ``yesterday'' at current position $k$ (blue line). We compare the activity level with the average activity of the participant in competitions regardless of his/her position (green line). The difference between the lines can be interpreted as the ``boost'' participant at current rank $k$ gets because of their current position.

We observe that yesterday's leader always increases their activity on the next day beyond their average in-competition activity. Similarly, second placed person only slightly increases their activity. Participants at current positions 3 and 4 maintain their level of activity to be the same as their average in-competition activity. However, participant at rank 5 (last position) performs worse than expected. This is interesting as it seems to suggest that while today's leader keeps his/her next day activity above the baseline, the person at the last position drops their activity to the level even below their baseline.

%This figure clearly illustrates that users with cumulative rank 1 (they have done more steps in the previous days than all the other participants) perform better the next day than their average number of in-challenge steps on that day. This is while the users with cumulative rank 2, 3, and 4 perform almost on par with their average and the last guy always performs worse as if he is discouraged.

\section{What Makes A Competition\\Engaging?}
% !TEX root = paper-competition.tex

%prob winning for cumulative leader
%final fig 11,12,13

%\jure{y-axis ideas: activity of the last person, relative increse of the last person, avg. increase of participants}

%\begin{figure}[th!]
%  \centering
%  \includegraphics[width=0.5\columnwidth]{FIG/WWW_cumrank_winner.pdf}
%  \captionof{figure}{Probability of the users with cumulative rank 1 on each day to be the eventual winner of the challenge}
%  \label{fig:WWW_cumrank_winner}
%\end{figure}

The focus of the last section was on the winner of the competition. In this section, we look at the dynamics of the competition more broadly. In particular, we  focus on more engaging competitions in which there is a close race for the top position with participants changing rank position multiple times. We will observe that these competitions are more successful in increasing the overall activity of the participants.  We will also identify the impact of other important factors such as the gender composition of the group as well as the inherent competitiveness of the participants. 

%are more successful in increasing the level of efforts of the participants? Is there a close race for the top position with participants changing rank multiple times? 

\xhdr{Probability of winning}
We start with the simplest possible model of competition as our null hypothesis. Consider the model in which all participants increase their activity level uniformly and at the same rate. In other words, they  start with their baseline levels of activity, then increase it to the in-competition level, and then effectively keep that same activity level throughout the competition. If that were the case, then we would expect that the person who takes the most steps after the first day of the competition would finally also win the competition.

\begin{figure}[th!]
  \centering
  \begin{subfigure}[b]{0.49\columnwidth}
      \includegraphics[width=\textwidth]{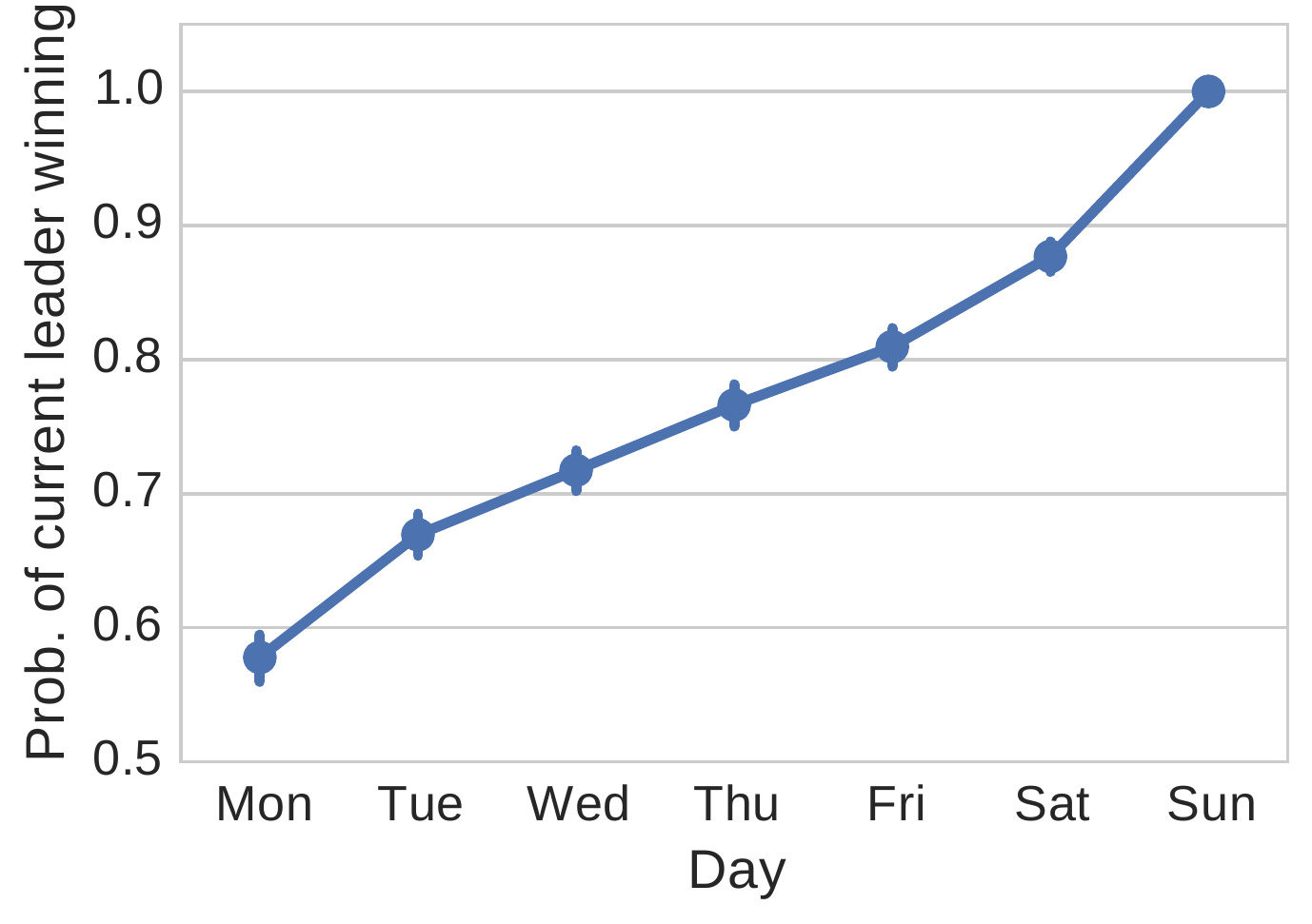}
      \caption{Prob. of winning}
      \label{fig:WWW_cumrank_winner}
  \end{subfigure}
  \begin{subfigure}[b]{0.49\columnwidth}
      \includegraphics[width=\textwidth]{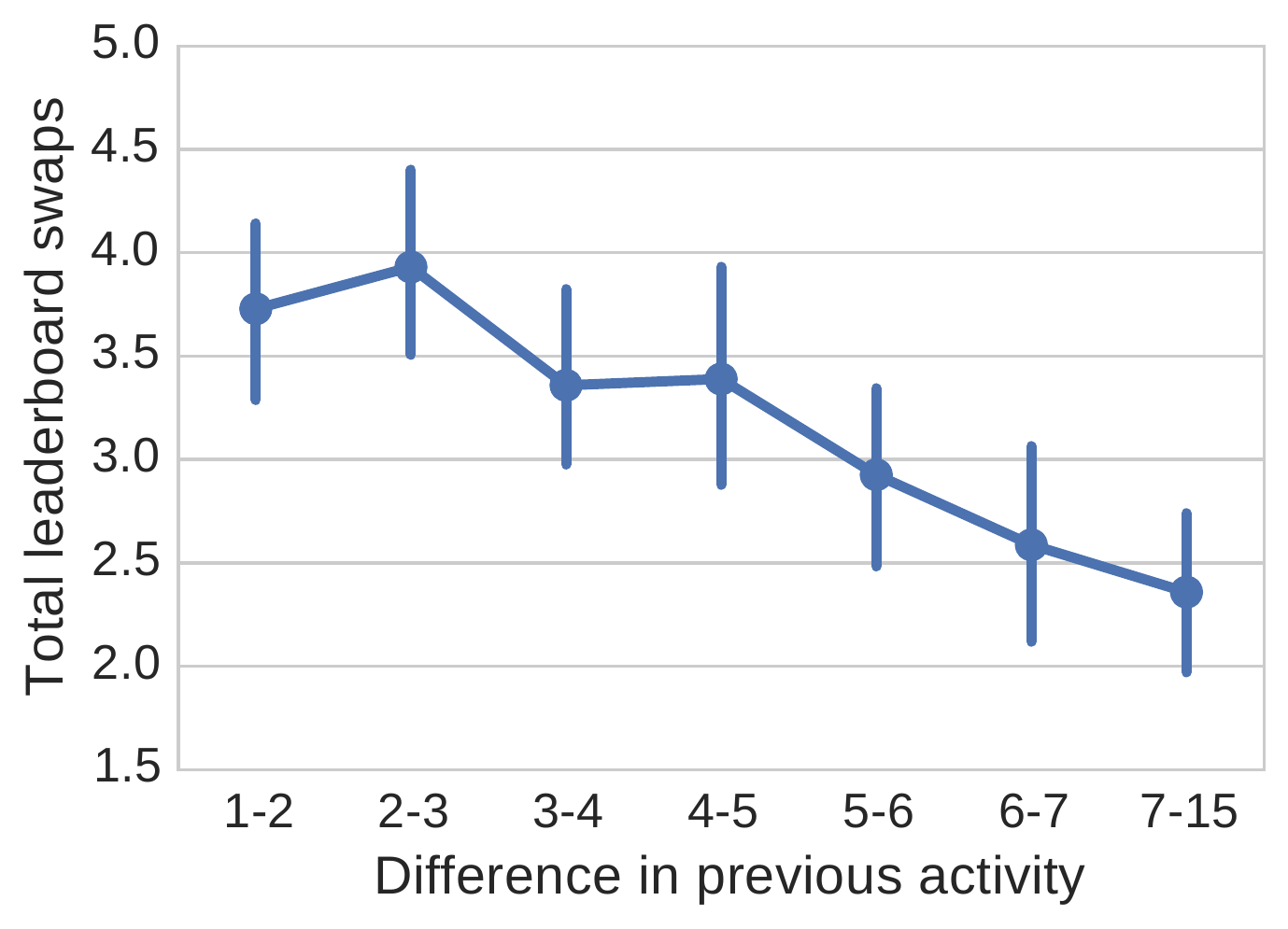}
      \caption{Leaderboard swaps}
      \label{fig:WWW-swaps_vs_max-min}
  \end{subfigure}
  \begin{subfigure}[b]{0.49\columnwidth}
      \includegraphics[width=\textwidth]{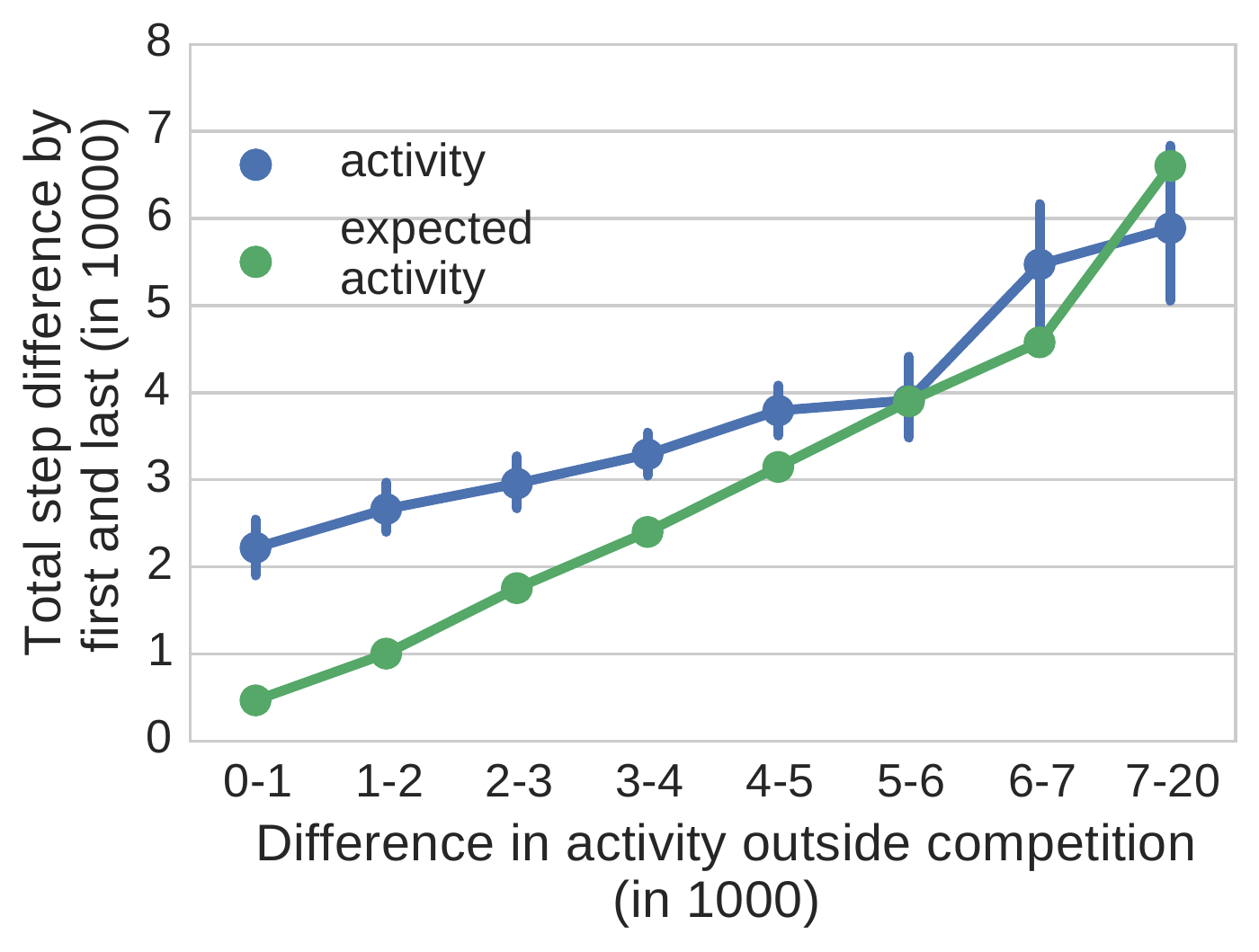}
      \caption{Activity spread}
      \label{fig:WWW-totalabs_vs_expectedabs}
  \end{subfigure}
  \begin{subfigure}[b]{0.49\columnwidth}
      \includegraphics[width=\textwidth]{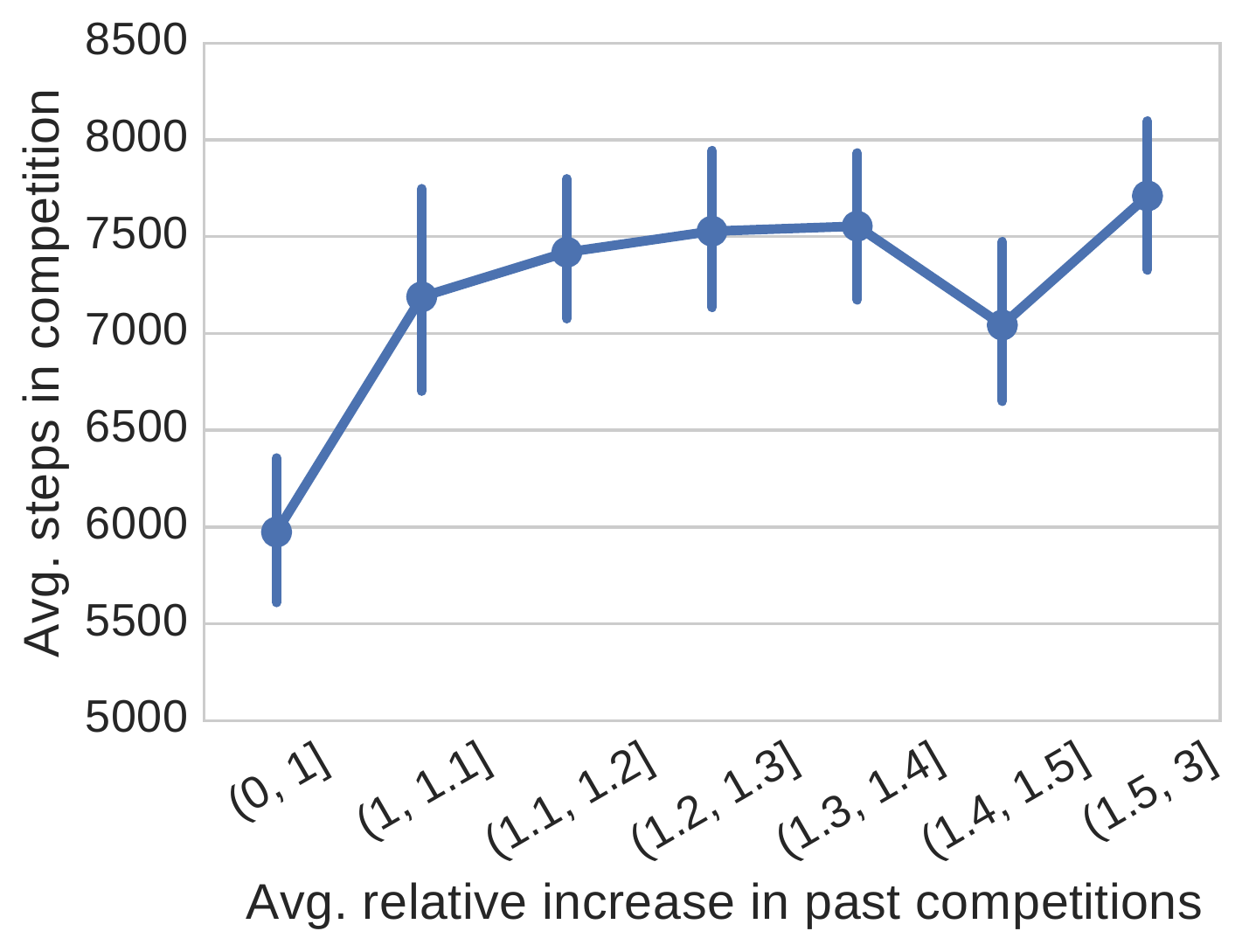}
      \caption{Activity increase}
      %\jure{Redo x-axis so that green is a diagonal line} \jure{Plot x-axis: increase in the LAST competition}
      \label{fig:WWW-deltaactivityrel_vs_reldeltamean}
  \end{subfigure}
  %\begin{subfigure}[b]{0.49\columnwidth}
  %    \includegraphics[width=\textwidth]{FIG/WWW-relchangevsmean.pdf}
  %    \caption{Relative change}
  %    \label{fig:WWW-relchangevsmean}
  %\end{subfigure}
  \caption{Competition dynamics. (a) Probability of current leader winning as a function of time; (b) Number of leaderboard swaps as a function of the difference between the most and the least active participant; (c) Total step difference between the top and the bottom ranked participant at the end of the competition as a function of the difference between the most and the least active participant before the competition. (d) Average participant's steps per day in a competition vs. average relative increase in past competitions.}
  \label{fig:}
\end{figure}

We examine this hypothesis in Figure~\ref{fig:WWW_cumrank_winner}, where we quantify the probability that the leader on a given day finally wins the competition.  Again, we examine 7 day competitions which all start on a Monday and end on a Sunday. We measure how often the current leader wins the overall competition. We observe that the leader after day 1 of the competition tends to win 58\% of the cases. The probability then linearly increases up to 1.0, meaning that the current leader on day 7 always wins the competition (because the competition is finished).

So overall, in slightly more than half of the competitions, the early leader can maintain the lead throughout the competition. Next, we will look at the dynamics of the more interesting competitions in which there are multiple changes in the leaderboards position and the  participants truly compete with each other.  

\xhdr{The dynamics of close competitions}
One way to quantify the competitiveness of a race is to measure the number of leaderboard changes (or swaps) as the competition unfolds. Here, Figure~\ref{fig:WWW-swaps_vs_max-min} plots the number of leaderboard changes as a function of the absolute difference between the most and the least active participant in the period before the competition. In other words, the figure shows how inequality in the baseline level of activity of participants affects the number of leaderboard swaps. We observe that when the difference in baseline activity is small, there are about 4 leaderboard changes in a competition. However, when this difference increases to 7-15 thousand steps per day, then the number of leaderboard changes drops but still remains non-trivial at around 2.5.

%\jure{What exactly are we measuring here? swaps or changes, only leader or any change. How big competitions, how long of a duration. Basically, the numbers seem a bit small, which could be due to short competitions with few participants.}

Another way to quantify how competitive a competition is to measure the final total difference in the number of steps between the top and the bottom placed participants (Figure~\ref{fig:WWW-totalabs_vs_expectedabs}). We plot the relationship between the final step difference between the winner and the loser of the competition as a function of the absolute difference between the most and the least active participants in the period before the competition (blue line). 
The green line provides a null-model that quantifies the expected final difference. Here we simply take the baseline difference and multiply it with the  duration of the competition. Our reasoning is that if, for example, in the baseline period the daily activity difference between the most and the least active person is say 1,000 steps, then the expected final difference after a 7 day competition would be 7,000 steps.

Examining Figure~\ref{fig:WWW-totalabs_vs_expectedabs} we make two  observations. First, as the inequality of the baseline physical activity of the participants increases, the final-step difference also increases. Second, we observe that for inequalities of less than 5,000 steps per day, the final difference is in fact larger than what would be expected under the null-model. This means that in tight competitions the winner strongly increases their activity level and the difference in activity grows larger. We also observe that when participants with very different baseline levels of activity compete, the final difference is in fact smaller than the baseline. This means that when uneven people are matched to compete, their level of activity actually gets closer to each other and the effect of the competition on the physical activity is smaller.

\xhdr{Competitiveness of participants} Lastly, we also examine how the past increases in the activity of participants determines their overall increase in the current competition. What happens if a competition is comprised of competitive participants -- the ones who had increased their activity levels significantly in previous competitions? Do they increase the level of activity of the whole group? Do these competitive tendencies have a compounding effect and raise the overall activity level significantly? 

We perform the following experiment. For every competition we compute the relative increase in activity in past competitions averaged over all participants. This gives us a sense of how competitive the participants are in the competition.  We then also compute the average activity of participants in the current competition to understand how the competitiveness of participants affects the overall activity.

%First, we define competitiveness $c_p$ of participant $p$ to be the relative increase of participant $p$ in competition over their baseline activity. Then for every competition $C$ with participants $P$, we compute two quantities: (1) The average competitiveness of people in $P$ but in the past competitions. This way we quantify how overall competitive is the set of participants $P$. And (2), for the same set of participants $P$ we also compute their competitiveness (average relative increase in activity) in the current competition $C$. As a null-model we also plot the expected average relative increase in activity if every person in $P$ would simply contribute their average historic competitiveness.

Figure~\ref{fig:WWW-deltaactivityrel_vs_reldeltamean} shows the results. We observe that as the average historic competitiveness of participants increases so does the overall competitiveness of the competition. In other words, if we put together people who tend to increase their activity by a lot, then they will also increase it in the current competition.  Not surprisingly, we also observe that  when competitions are comprised of people who do not tend to increase their activity in competitions then the average activity is low.

More interestingly, the effect of the competitiveness of the participants on the average level of activity is sharp but then it quickly levels off.  In particular, the overall activity tends to stabilize after the average past increase of activity (of participants when in competition) is over 10\%. This means that as soon as competitions are comprised of participants who generally tend to increase their activity during competitions then the average steps per participant steps will reach up to 7,500 per day. If the competition is among  extremely competitive participants, the average activity only slightly increases above that level.

\begin{figure}[t!]
\centering
%\begin{minipage}{.47\columnwidth}
  \centering
  \includegraphics[width=0.75\columnwidth]{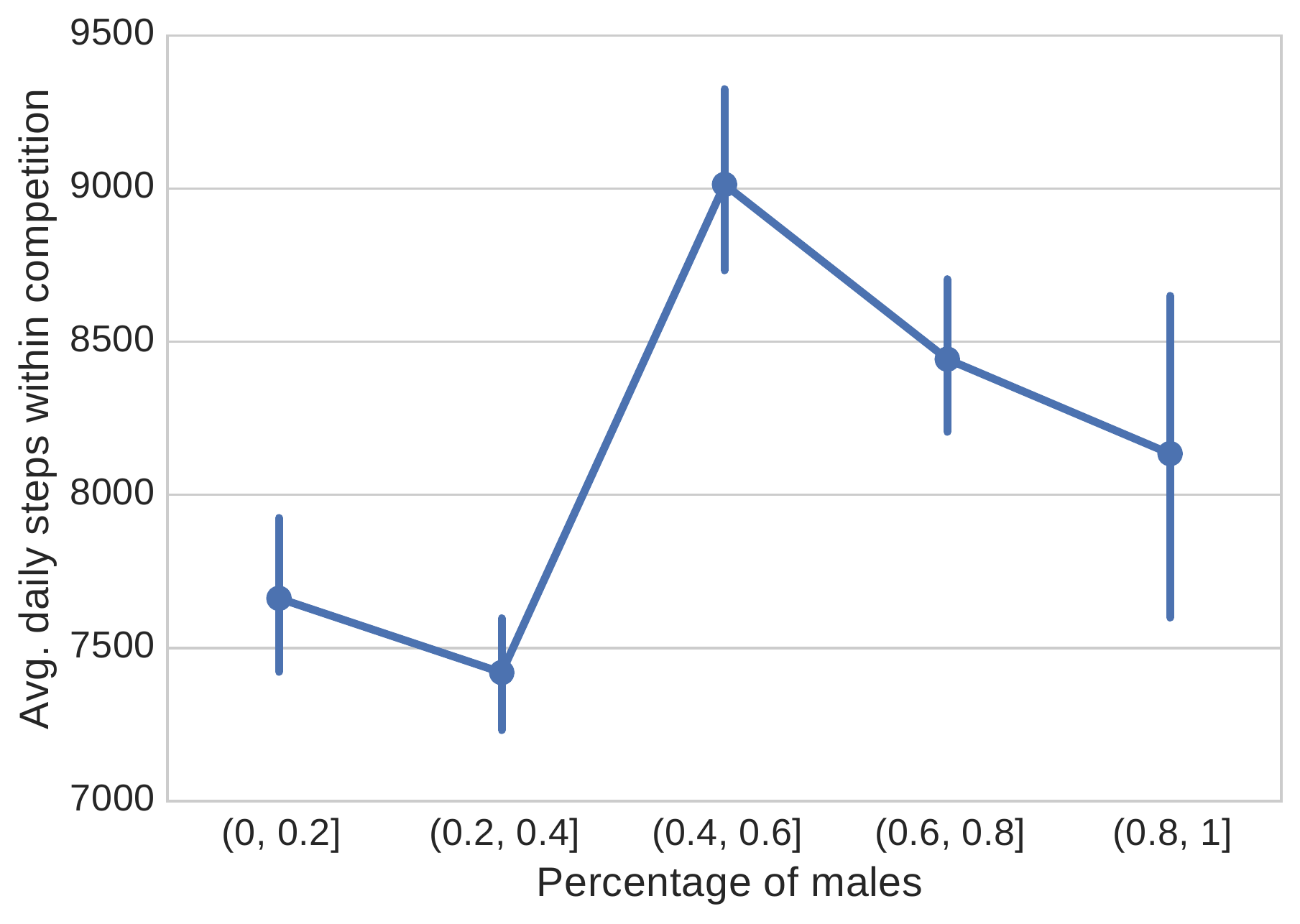}
  %\includegraphics[width=0.35\columnwidth]{FIG/WWW-stepsperday.pdf}
%   \vspace{-2mm}
  \captionof{figure}{The average number of daily steps taken by participants as a function of the fraction of male competitors.}
%   \vspace{-3mm}
%\end{minipage}%
  \label{fig:WWW-steps_vs_malepercentage}
\end{figure}

\xhdr{The effect of gender diversity}  Figure \ref{fig:WWW-steps_vs_malepercentage} plots the average number of steps taken by participants in the competition by across varying compositions of male and female participants in the competition. Note that competitions with a balanced number of female and male participants have the highest level of activity. This is surprising because men are on average more active than women (\ie, they take more steps per day) both in our dataset as well as published estimates~\cite{Bassett2010}. Nevertheless, competitions with all male participants have a smaller average than competitions in which the number of men and women is close to each other. This observation suggests that gender diversity can lead to more active and beneficial competitions.

%\jure{We observe that less competitive people increase their activity more than the more competitive people. We could explain this by more competitive people getting tired.}

%in the current competition participant's level of activity is the same as in the past. We observe that people who in the previous competitions increased their for less than 10\%, tend to increase it more in the current competition. On the other hand, participants who in the past competitions increase more than 20\%, tend to slightly drop their increase in activity. This means that high increases in activity generally cannot be sustained by the participants and that a low performance tends to be followed by a high performance competition and vice versa.

\xhdr{Summary} We observe that the composition of participants greatly affects the dynamics of the competition. In particular, if highly unequal participants get matched to compete, then the competition suffers and the overall effect of the competition on the physical activity drops significantly. Second, competitions with a balanced gender ratio are more effective in increasing the level of activities. Lastly, the effect of  competitiveness of participants as defined by their performance in the previous competitions has a sharp impact on the level of activity but its impact quickly levels off.  With these insights in mind, the next section takes on the task of predicting such dynamics to serve as a guideline for the formation of interesting and engaging competitions.

%\amin{I removed this and replaced it w the last paragraph:} With these insights in mind, we shall next  examine the question of competition formation: How shall one select competition participants so that the competition has the optimal effect on the physical activity of the participants? 
% For figure 8C, again the blue line is the average of relative increase of steps for each user inside challenge (comparing to his outside challenge activity in the past) and the green line is the mean of each bin

\section{Predicting Engaging Challenges}
% !TEX root = paper-competition.tex

%\tim{if we have time, try top quartile vs bottom quartile. will focus on related work and writing first.}

% \jure{Tim, would it be possible to add some more results to this section. Right now it is very short and very light. We have about 1 page to spend on this section. Could you also compare how different types (groups) of features perform -- create groups of features and then do an ablation study to see which groups of features are most predictive in which task. We really need this in order to make the paper feature complete.}

\enlargethispage{\baselineskip} % Tim: 2017/01/08 getting to 8 pages

In the previous sections we studied the factors of engaging competitions -- those that lead to large increases of activity, close races between the first and the last ranked user, and competitions with many changes in the leaderboard ranking.  
This section builds on those insights and builds a series of models to predict which competition will be particularly engaging given only information available before the competition begins. We demonstrate that while there is large variability in the engagement of competitions, the factors described in this work allow to predict competition engagement with significant accuracy.

\begin{table}[t]
% \vspace{-3mm}
\centering
% \small % \footnotesize
\resizebox{.999\columnwidth}{!}{%
\begin{tabular}{lrrr}
  %\hline
  % \newline
  \toprule
    \textbf{Model} &  \textbf{$\Delta$ACTIVITY} &  \textbf{FIRST-} &  \textbf{RANK} \\ %\hline
     & & \textbf{LAST} &  \textbf{SWAPS} \\
  \midrule
   Random & 0.500 & 0.500 & 0.500 \\
   Logistic Regression & 0.698 & 0.749 & 0.643 \\ %  ($L_1$ reg.)
   Gradient Boosted Trees & 0.719 & 0.742 & 0.611 \\
  \bottomrule
 \end{tabular}
 }
% \vspace{-1mm}
 \caption{
 Prediction performance of several models predicting three outcomes of engaging competitions. 
 Performance values correspond using the area under the ROC curve. 
 }
 % \vspace{-2mm}
 \label{tab:prediction}
 \end{table}

\begin{table}[t]
%\vspace{-2mm}
\centering
% \small % \footnotesize
\resizebox{.999\columnwidth}{!}{%
\begin{tabular}{rlrrr}
  %\hline
  % \newline
     &\textbf{Model} &  \textbf{$\Delta$ACTIVITY} &  \textbf{FIRST-} &  \textbf{RANK} \\ %\hline
     & & & \textbf{LAST} &  \textbf{SWAPS} \\
  \midrule
   1 & None & 0.500 & 0.500 & 0.500 \\
   2 & \# participants & 0.507 & 0.682 & 0.500 \\
   3 & User demographics & 0.645 & 0.668 & 0.622 \\
   4 & User experience & 0.584 & 0.689 & 0.599 \\
   5 & Prev. activity (outside) & 0.668 & 0.707 & 0.609 \\
   6 & Prev. activity (during) & 0.576 & 0.751 & 0.606 \\
   7 & Prev. increases (during) & 0.717 & 0.691 & 0.641 \\
   8 & All features & 0.719 & 0.742 & 0.611 \\
  \bottomrule
 \end{tabular}
 }
% \vspace{-1mm}
 \caption{
 Prediction performance of Gradient Boosted Tree models using different feature sets predicting three outcomes of engaging competitions. 
 Performance values correspond using the area under the ROC curve. 
 }
% \vspace{-2mm}
 \label{tab:prediction_features}
 \end{table}

\xhdr{Predicted outcomes}
%\jure{I don't really like the all caps ``$\Delta$ACTIVITY'', makes the text look very funny.}
We formulate the prediction task as a binary prediction of whether or not a competition will be engaging based on three different outcomes:
\begin{itemize}%[noitemsep]
  \item $\Delta$ACTIVITY: The average relative increase of activity during the competition compared to activity before the competition across all participating users. For the baseline activity before the competition we exclude any activity that happened during previous competitions. Ideal engaging competitions would lead to large increases in activity across all users.
  \item FIRST-LAST: The absolute difference in total number of steps between the first and the last ranked user over the 7 day competition. Engaging competitions are close races where the first and last user are not too far apart.
  \item RANK SWAPS: The number of total changes in the leaderboard over the time of the competition measured as the minimum number of inversions in the day-to-day rankings. Engaging competitions are competitions where participants are competing in a tight race leading to many changes in the leaderboard over the course of the competition.
\end{itemize}
Each of these outcomes is a continuous variable which we transform into a (approximately) balanced binary prediction problem by splitting at the median value ($\Delta$ACTIVITY 1.199 factor of increase, FIRST-LAST 37,421 total steps, RANK SWAPS 4 swaps). 

\xhdr{Data and methods}
We use the dataset of 7 day competitions and at least three participants (N=2,432).%2,262 after trimming.
% \tim{apparently this number is smaller than Table 1 due to outliers? We should use the same dataset everywhere. Please update Table 1 accordingly.}
We use 75\% for training and 25\% for testing at random.
Area under the ROC curve is used as a measure of predictive performance on the test set.
We report performance for Gradient Boosted Tree and $L_1$ penalized logistic regression models and optimize number of trees, tree depth, learning rate, and regularization parameter through 5-fold cross-validation on the training data.

\enlargethispage{\baselineskip} % Tim: 2017/01/08 getting to 8 pages

\xhdr{Features used for learning}
We featurize the state before the competition as follows. In all cases we only use data from before the start of the competition:
% If features are missing we impute zero and include a binary variable indicating missingness.
\begin{itemize}%[noitemsep] %,nolistsep
  % \item Competition type: We use the duration (3, 5, or 7 days) and the number of participants in the competition.
  \item {\bf Participants:} Number of participants in competition.
  \item {\bf User demographics:} We use the number and fraction of male users and the number and fraction of obese users (BMI>30) as well as the median age of all participants.
  \item {\bf User experience:} We use number and fraction of users who have participated in a competition before and the number and fraction of users who have won a competition before. 
  Furtermore we use summary statistics (further explained below) of the distribution over how many times users have participated in a competition or won a competition. 
  \item {\bf Previous outside-competition activity:} We use summary statistics of the distribution over number of steps taken in previously outside of any competitions by all the participants.
  \item {\bf Previous in-competition activity:} We use summary statistics of the distribution over number of steps taken in previous competitions by all the participants.
  \item {\bf Previous increases in competition activity:} We measure how much each user has increased their daily number of steps in previous competitions (relatively and absolutely). We then use summary statistics over this distribution of increases across all participants.
\end{itemize}

As summary statistics we use mean, standard deviation, max, min, second largest, second smallest, difference between max and min, difference between max and second largest, difference between second smallest and min, and the number/fraction of users within 0.5, 1.0, and 2.0 standard deviations of the mean.

\xhdr{Predictive performance} 
Prediction accuracies using the ROC AUC measure are shown in Table~\ref{tab:prediction}.
Overall, we observe encouraging predictive performance.
We find that Logistic Regression models and Gradient Boosted Trees perform similarly well after optimization of parameters through cross validation.
Predicting $\Delta$ACTIVITY yields a performance of 0.72  ROC AUC.
The difference between the first and last user in total steps yields 0.75  ROC AUC.
Predicting whether the competition will see many leaderboard swaps is more challenging and the model achieves 0.64 ROC AUC.

Overall, we show that it is able to predict which competition will be engaging before it start with ROC AUC scores between 0.64-0.75.

\xhdr{Factors affecting predictability}
Next, we study which factors are particularly helpful in predicting the three outcomes.
This is relevant to application settings where not all features might be available (\eg, missing knowledge about participants previous activity level, previous in-competition activity, or missing information on user demographics).
We investigate performance of the individual feature sets described above.
We report prediction accuracies for Gradient Boost\-ed Tree models as they performed slightly better than than Logistic Regression models, particularly on small sets of features.
The predictive performance for the individual feature sets is reported in Table~\ref{tab:prediction_features}.

We observe that knowing the number of participants (Model 2) is only useful for predicting the final difference between the first and last ranked users but not for the other two measures. This is likely due to the effects observed in Figure~\ref{fig:WWW-steps_per_counts_comp} which showed that the difference between the first and last ranked users increases in larger competitions.
Knowing just basic user demographics (age, gender, and obesity status) allows to predict all three outcomes with significant accuracy (Model 3). In particular, the number of RANK SWAPS can be predicted with ROC AUC of 0.622 which is as good as using all features (Table~\ref{tab:prediction}).
%\tim{we could try figure for higher fraction male - more swaps? there are also references for competition and gender effects. e.g. males enjoy it more or so}

Past experience with competitions of all users (Model 4) is also predictive of engaging competitions. In particular, it allows good predictive accuracies for predicting FIRST-LAST (0.689 ROC AUC) and relatively good performance for predicting RANK SWAPS (0.599 ROC AUC) close to the the best performing model.% (Model 7).

Knowing the exact previous activity levels of the users -- both outside (Model 5) and inside of competitions (Model 6) -- performs even better. Outside of competition activity is a good indicator for the final difference between the first and last ranked users (Figure~\ref{fig:WWW-totalabs_vs_expectedabs}) with a prediction performance of 70.7\% ROC AUC. 
Prediction of FIRST-LAST improves even more when knowing previous inside competition activity levels with 75.1\% ROC AUC (best performing model for FIRST-LAST).

Baseline activity levels outside of competitions and activity levels during previous competitions allows to calculate how much each user increased (or decreased) their activity in response to previous competitions (Model 7).
This allows for high predictive performance for both $\Delta$ACTIVITY (0.717 ROC AUC; close to best performing model) and RANK SWAPS (0.641 ROC AUC; best performing model).

As expected, using all combined features allow for high predictive performance (Model 8). However, it is interesting to not that for FIRST-LAST and RANK-SWAPS the model performs slightly worse compared to just using the best subset of features.

\xhdr{Implications}
The prediction results are encouraging in multiple ways.
First, we can predict key outcomes capturing various aspects of engaging competitions with significant accuracy. 
% achieve decent accuracy levels across all three outcomes capturing various aspects of engaging competitions. 
The proposed models could be used when recommending which people to add to a competition to optimize the likelihood of high engagement.
Second, we demonstrate that having access to only small subsets of the features studied in this work still allows for models achieving similarly high accuracy.
This has direct implications for the use of competition features in practice. 
Many mobile apps and websites use, but not yet optimize, competitions to engage with their users. 
Given our results, they could use for example basic demographic information or outside competition activity levels to create particularly engaging competitions.
We also note that during competition behavior is not necessary for high predictive performance (\eg, Models 3 and 5 in Table~\ref{tab:prediction_features}). This partially alleviates potential cold start problems when creating new competition-based applications.

\section{Conclusion}
	% !TEX root = paper-competition.tex
%% ALI:
%The utmost important factor promoting physical health is exercise. Regular exercise can contribute to a healthy lifestyle and enhance a person's quality of life. In this paper we studied a large dataset gathered by a public health application measuring the total number of daily steps taken by users. This application has a built in social network which was incorporated into the application about a year ago, giving users the ability to make connection and commence competions with their firends. We studied the effect of competitions on user behavior and showed that in general, users increase their activity level when participating in a competition. We found there are many important factors contributing to an engaging competition such as age, gender, BMI, sexual diversity and prior activity levels of users. Furthermore, we integrated our results into a learning algorithm to predict the outcome of challenges which in turn helps us design evenly-matched competitions that are most likely to help users increase their daily activity.

% !TEX root = paper-competition.tex

% \enlargethispage{\baselineskip}
 
The focus of this paper is on how competition incentivizes users to increase their physical activity levels. We analyzed a large dataset gathered by a smartphone activity tracking app measuring the total number of daily steps taken by a user. We studied games where multiple users compete over several days with the goal of achieving the highest total number of steps. 

We measured the effect of competitions on user behavior and showed that in general, users increase their activity level by 23\% when participating in a competition. 
We also studied the design elements contributing to an engaging competition and identified the importance of factors like age, gender, BMI, diversity, and prior activity levels. 
We found that the group compositions strongly affects the dynamics of the competition, which leads to important design implications for exergames and mobile health applications:   
%In particular, if highly unequal participants get matched to compete, then the overall dynamics of the competition suffers and the overall effect of the competition on the physical activity drops significantly. Furthermore, competitions with a balanced mix of both men and women are more effective in increasing activity.
% \todo{
\begin{enumerate}
  \item Competitions lead to increases in physical activity and constitute a viable design element able to reach a broad user base across a wide variety of user demographics. 
  \item Competing participants should have similar pre-com\-peti\-tion activity levels. Otherwise the effect of the competition on physical activity drops significantly.
  \item Competitions should have a balanced mix of both men and women.
  \item Competitions should ideally include some participants who have previously increased their activity in response to competitions to encourage the other participants.
\end{enumerate}
% } %end todo 
Finally, we leveraged these insights in a statistical model that predicts how effective and engaging a given competition is going to be. Such models can lead to designing more evenly-matched competitions that are most likely to help users increase their daily activity.
In summary, this work enhances the understanding of effective mechanisms that can potentially be used to engage people in healthier behaviors. % Results of this work can be applied across a variety of mobile health apps to recommend evenly-matched competitions to users which are most likely to help them become more active.

\xhdr{Acknowledgments}
We thank Azumio for providing us with ano\-nymized activity data that enabled this study.
This research has been supported in part by NSF IIS-1149837, NIH BD2K Mobilize Center, ARO MURI, DARPA SIMPLEX
and NGS2, Stanford Data Science Initiative, Lightspeed,
SAP, Chan Zuckerberg Biohub, and Tencent.

% % \pagebreak
% \balance
% \bibliographystyle{abbrv}
% \bibliography{refs}

% \clearpage
% \pagebreak
\balance

% no space between bib items: http://tex.stackexchange.com/questions/17360/reduce-spacing-in-bibliography-using-biblatex
\let\oldbibliography\thebibliography
\renewcommand{\thebibliography}[1]{\oldbibliography{#1}
\setlength{\itemsep}{0pt}} %Reducing spacing in the bibliography.

% \small
\let\secfnt\undefined
\newfont{\secfnt}{ptmb8t at 12pt}
\bibliographystyle{abbrv}
\bibliography{refs}

 %\appendix
 %\section{Appendix}
 %\label{sec:appendix}
 %\input{080appendix}

\end{document}